\documentclass{article}
\pdfoutput=1

\usepackage{arxiv}

\usepackage[utf8]{inputenc} 
\usepackage[T1]{fontenc}    
\usepackage{hyperref}       
\usepackage{url}            
\usepackage{booktabs}       
\usepackage{amsfonts}       
\usepackage{nicefrac}       
\usepackage{microtype}      
\usepackage{graphicx}
\usepackage{gensymb}
\title{Magnetic dilution by severe plastic deformation}

\author{
  Martin St\"uckler \\
  Erich Schmid Institute of Materials Science, Austrian Academy of Sciences\\
  Jahnstra{\ss}e 12, 8700 Leoben, Austria \\
  \texttt{martin.stueckler@oeaw.ac.at} 
   \And
  Lukas Weissitsch \\
  Erich Schmid Institute of Materials Science, Austrian Academy of Sciences\\
  Jahnstra{\ss}e 12, 8700 Leoben, Austria \\
   \And
   Stefan Wurster \\
     Erich Schmid Institute of Materials Science, Austrian Academy of Sciences\\
  Jahnstra{\ss}e 12, 8700 Leoben, Austria\\ 
   \And
  Peter Felfer \\
  Department of Materials Science, Friedrich-Alexander-Universit\"at Erlangen-N\"urnberg \\
  Martensstra{\ss}e 5, 91058 Erlangen, Germany 
   \And
  Heinz Krenn \\
  Institute of Physics, University of Graz\\
  Universit\"atsplatz 5, 8010 Graz, Austria 
  \And
  Reinhard Pippan\\
    Erich Schmid Institute of Materials Science, Austrian Academy of Sciences\\
  Jahnstra{\ss}e 12, 8700 Leoben, Austria 
  \And
  Andrea Bachmaier\\
  Erich Schmid Institute of Materials Science, Austrian Academy of Sciences\\
  Jahnstra{\ss}e 12, 8700 Leoben, Austria 
}

\begin{document}
\maketitle

\begin{abstract}
Mixtures of Fe and Cu powders are cold-compacted and subsequently deformed with severe plastic deformation by high-pressure torsion, leading to bulk samples. The dilution of Fe in the Cu matrix is investigated with SQUID-magnetometry, whereas the magnetic properties change as a function of Fe-content from a frustrated regime to a thermal activated behaviour. The magnetic properties are correlated with the microstructure, investigated by synchrotron X-ray diffraction and atom probe tomography. Annealing of the as-deformed states leads to demixing and grain growth, with the coercivity as a function of annealing temperature obeying the random anisotropy model. The presented results show that high-pressure torsion is a technique capable to affect the microstructure even on atomic length scales.
\end{abstract}


\section{Introduction}
Metastable phases exhibit a high potential in application, since attractive functional and magnetic properties were observed in such non-equilibrium states \cite{berkowitz1993giant, ambrose1993magnetic, he2002transition}. In particular, the immiscible Fe-Cu system is of great interest, as Fe, mixed with a low amount of Cu, exhibits desirable magnetostrictive properties \cite{gorria2004invar}. On the other hand, Cu mixed with low amounts of Fe shows the granular giant magnetoresistive effect \cite{xiao1992giant, wang1993investigation}. But all of the mentioned studies have in common, that sample preparation routes were used, which are hardly accessible for industrial applications in terms of bulk products. Whereas magnetron sputtering or vapour depositions turn out to be not feasible on large scales, upscaling is possible in the case of mechanical alloying or rapid solidification. Anyway, one has to prepare a bulky sample, either from powder or thin ribbons. Therefore, sample preparation directly in bulk form is desirable. Techniques of severe plastic deformation, such as high-pressure torsion (HPT), are capable to prepare such metastable phases directly in bulk form and can also be used at large scales. However the achievable degree of intermixing has to be examined. Therefore, non-equilibrium Fe-Cu samples are processed with HPT in this study, whereas intermixing is investigated for various Fe-contents with advanced microstructural characterization techniques in combination with SQUID-magnetometry.
\section{Experimental}
Powder mixtures, consisting of high purity powders (Fe: MaTeck 99.9\% -100 +200 mesh, Cu: AlfaAesar 99.9\% -170 +400 mesh), were prepared with three different Fe-contents, namely Fe07wt.\%-Cu, Fe14wt\%-Cu and Fe25wt.\%-Cu, and hydrostatically compacted at 5~GPa. The cylindrical samples (\textit{d}=8~mm, \textit{h}$\sim$0.5~mm) were subsequently deformed with HPT at 5~GPa for 100 turns at room temperature. The applied amount of shear strain at $r$=3~mm was $\gamma{\sim}$3000. More details on sample preparation can be found elsewhere \cite{stuckler2019magnetic}. All compositions were exposed to annealing treatments at 150\degree C, 250\degree C and 500\degree C for 1~h each, which were conducted in vacuum ($p{\leq}$10$^{-3}$~mbar) to prevent oxidation. Microstructural analysis was carried out with a scanning electron microscope (SEM; Zeiss LEO 1525), with attached electron backscatter diffraction (EBSD; Bruker Nano eFlash$^{\mathrm{F} \mathrm{S}}$), in samples tangential direction. Investigation on crystallographic compositions was carried out with synchrotron X-ray diffraction at DESY (beam energy: 100~kV; beam size: 0.2x0.2~mm$^2$). The beam was oriented in axial direction of the sample at $r$=3~mm. Further microstructural analysis was performed via atom probe tomography (APT) experiments on a LEAP4000~X at 50~K in laser mode with a pulse repetition rate of 250~kHz. Magnetic properties were investigated using a SQUID-magnetometer (Quantum Design MPMS-XL-7).
\section{Results and Discussion}
\subsection{Microstructural Characterization}
\begin{figure}
\includegraphics[width=\linewidth]{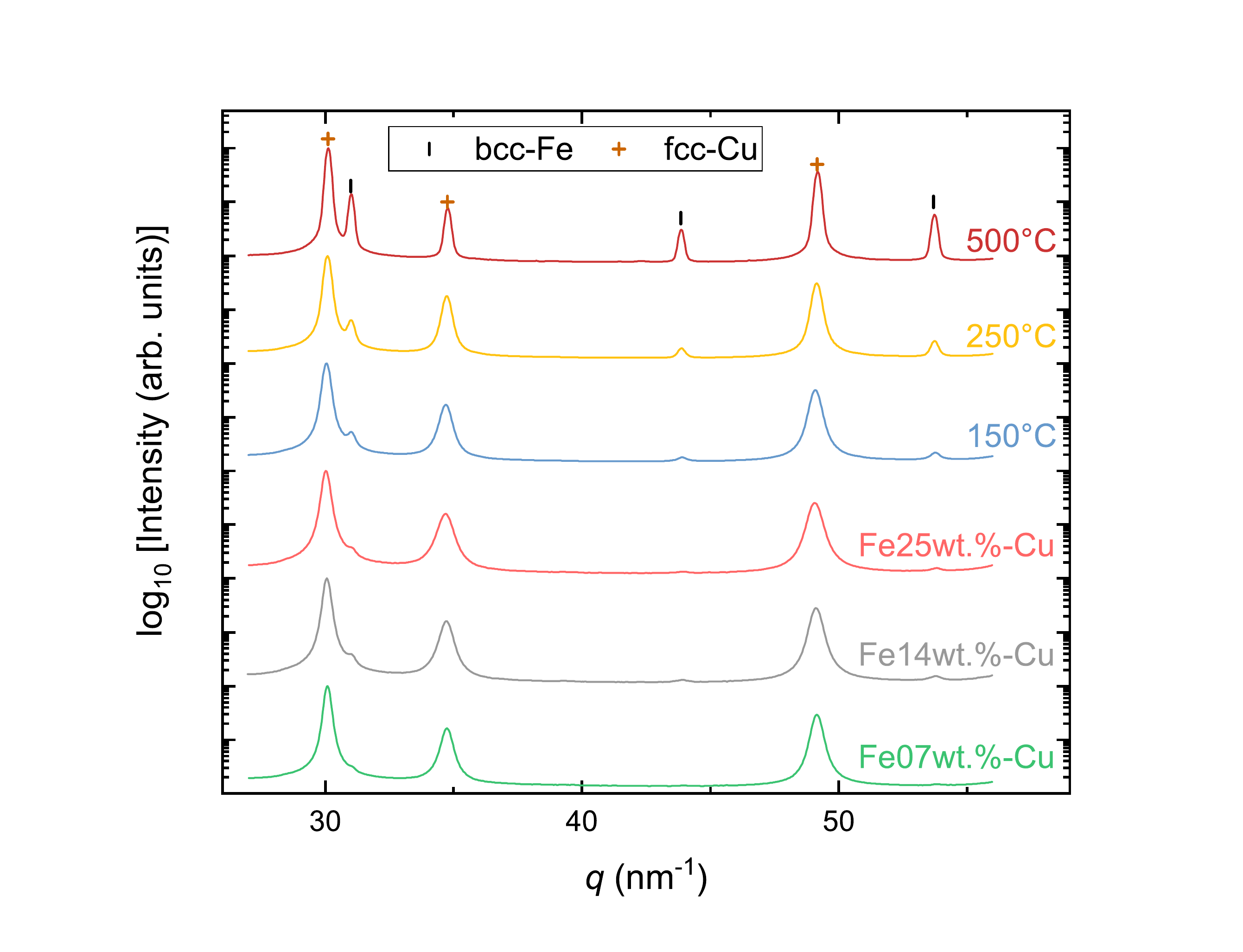}
\caption{Synchrotron-XRD diffraction patterns of as-deformed samples and of Fe14wt.\%-Cu in annealed states, measured in transmission mode.}
\label{fig:XRD}
\end{figure}

HPT-processing applies a shear strain onto the sample, which increases with increasing radius\cite{valiev2000bulk}, causing a microstructural evolution, which strongly affects the mechanical properties. To verify that a steady state is reached, Vickers hardness testing is carried out. The hardness shows a saturating behaviour at $r{\geq}$1~mm for 100 turns and therefore further investigations are carried out at large radii.
Fig.~\ref{fig:XRD} shows synchrotron XRD-patterns of as-deformed samples, as well as the patterns of Fe14wt.\%-Cu in annealed states as an example. The as-deformed samples exhibit very prominent fcc-Cu peaks, but tiny bcc-Fe peaks can also be identified. The latter are two orders in magnitude smaller and reveal the presence of residual Fe-particles. Annealed samples show an increasing intensity for bcc-Fe, indicating demixing. While the peak-width of the 150\degree C-annealed sample does not change with respect to the as-deformed state, the 250\degree C- and the 500\degree C-annealed sample both show a narrowing of peak-width arising from grain growth. Annealed Fe07wt.\%-Cu and Fe25wt.\%-Cu samples exhibit the same behaviour.
\subsection{Magnetic Properties}
\begin{figure}
\includegraphics[width=\linewidth]{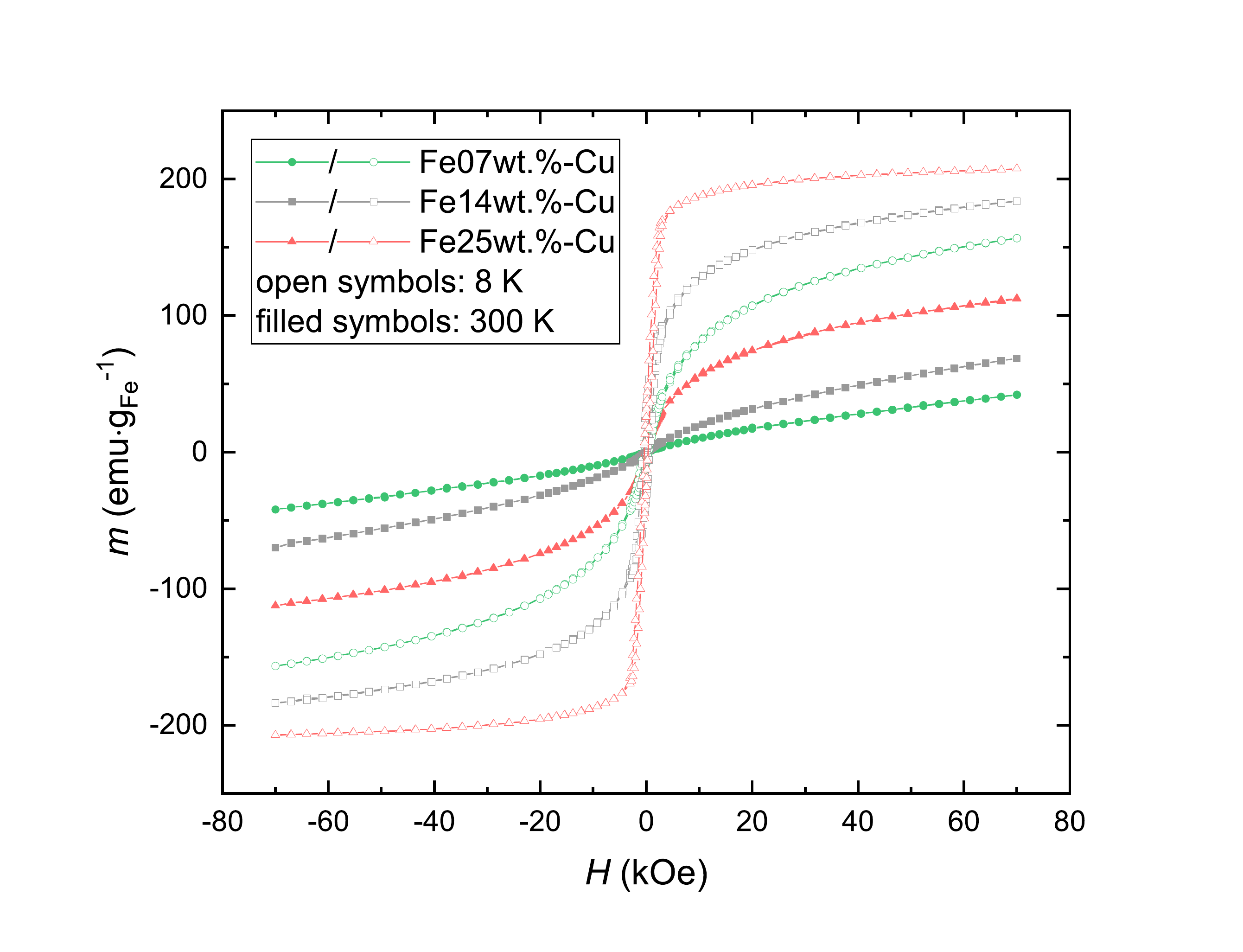}
\caption{Specific magnetization $m$ versus the applied field $H$ as a function of Fe-concentration, measured at 300~K and 8~K. Note: the saturation magnetization of bulk Fe is 220~emu/g.}
\label{fig:hyst_conc}
\end{figure}
\begin{figure}
\includegraphics[width=\linewidth]{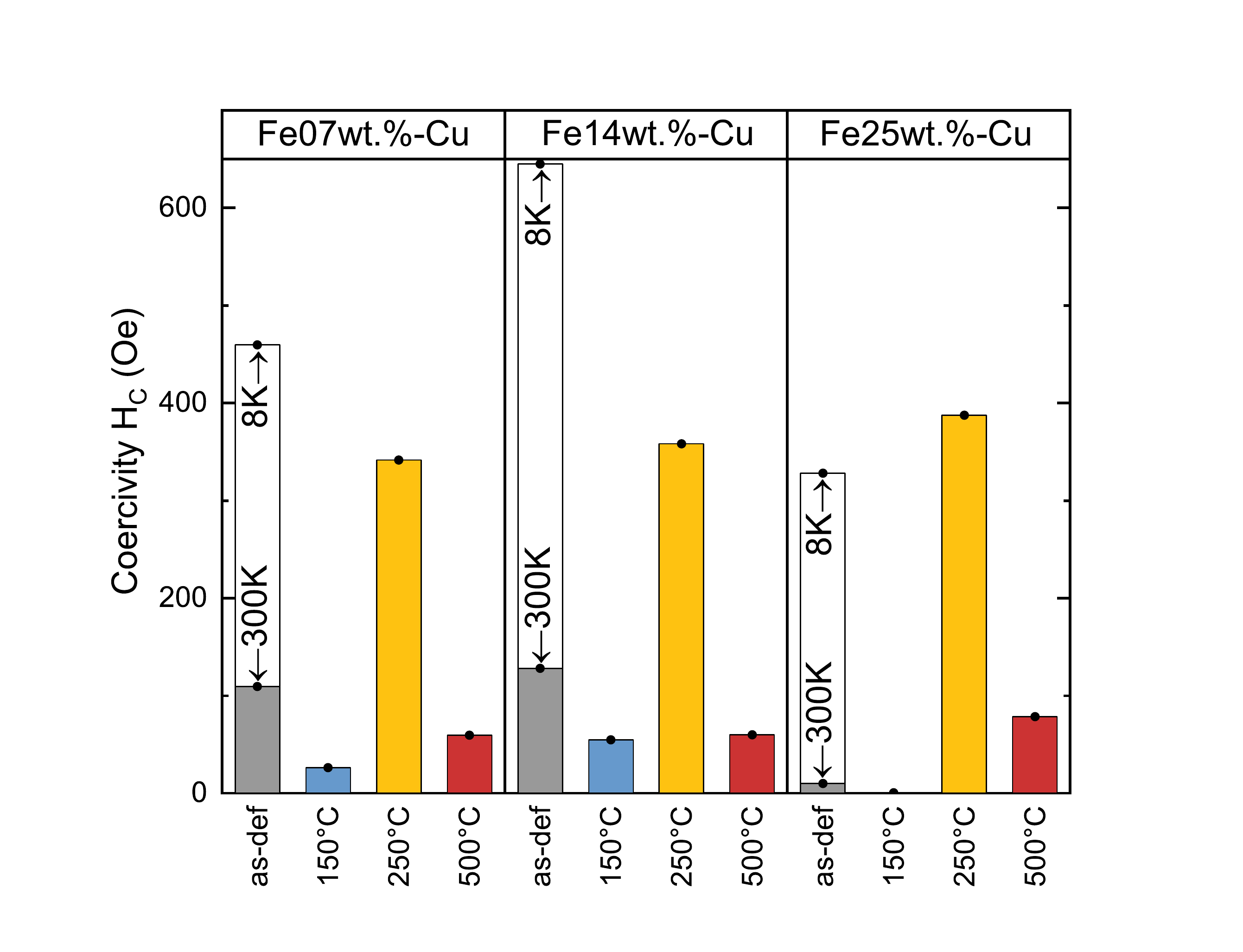}
\caption{Coercivity $H_C$ as a function of Fe-concentration and annealing temperature. The coercivity shows the same behaviour in all investigated compositions. Filled bars represent measurements at 300~K, whereas open bars represent measurements at 8~K. For the 150\degree C annealed state of Fe25wt.\%-Cu the coercivity is below the measurements resolution.} 
\label{fig:coercivity}
\end{figure}
\begin{figure}
\includegraphics[width=\linewidth]{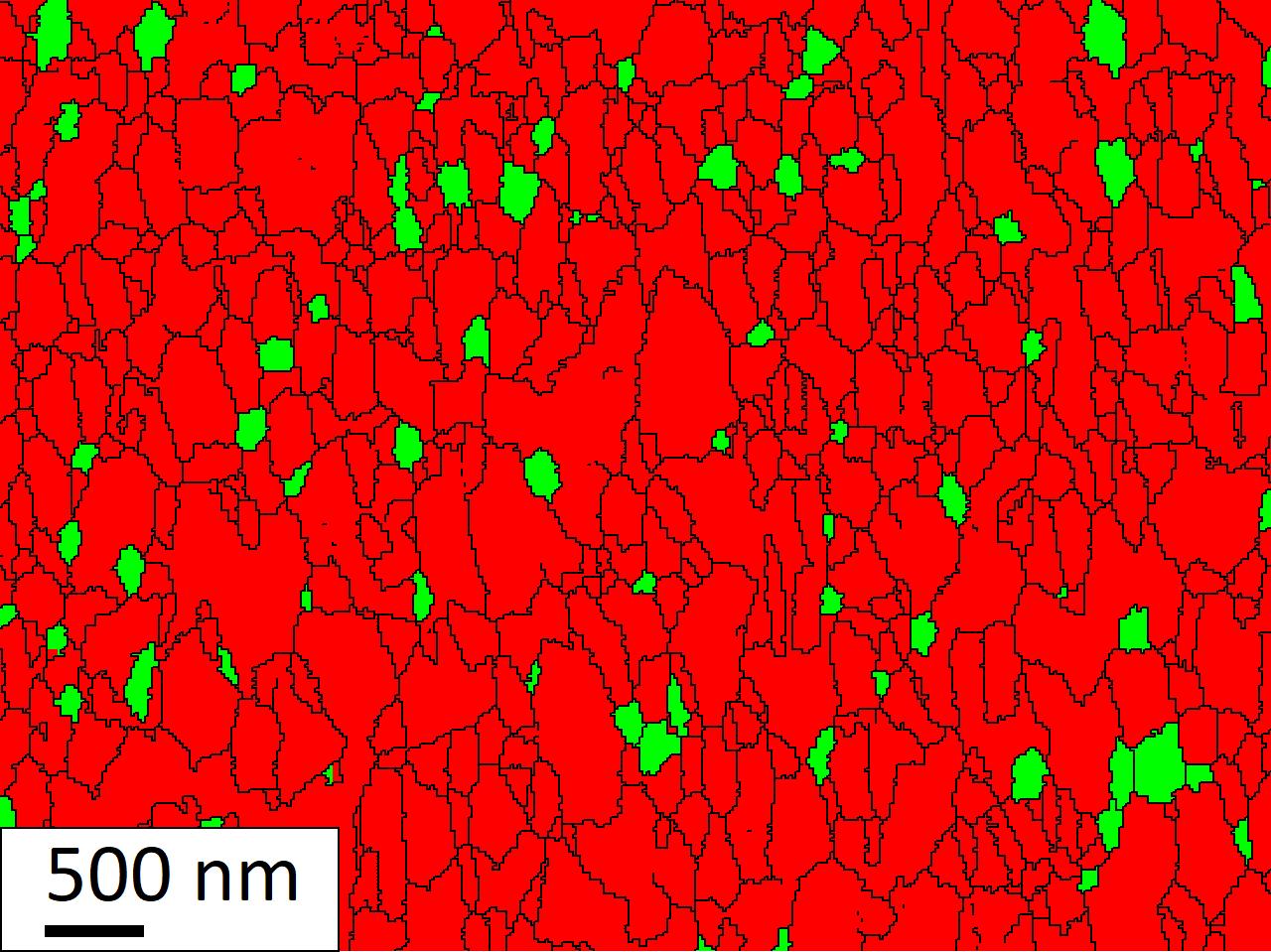}
\caption{Post-processed EBSD scan of Fe14wt.\%-Cu, annealed at 500$^{\circ}$C. Red regions represent Cu grains, while green regions represent Fe grains. Lines mark grain boundaries. The mean grain size of Fe is determined to (210$\pm$80)~nm. The shearing direction is parallel to the horizontal axis.}
\label{fig:ebsd}
\end{figure}
\begin{figure}
\includegraphics[width=\linewidth]{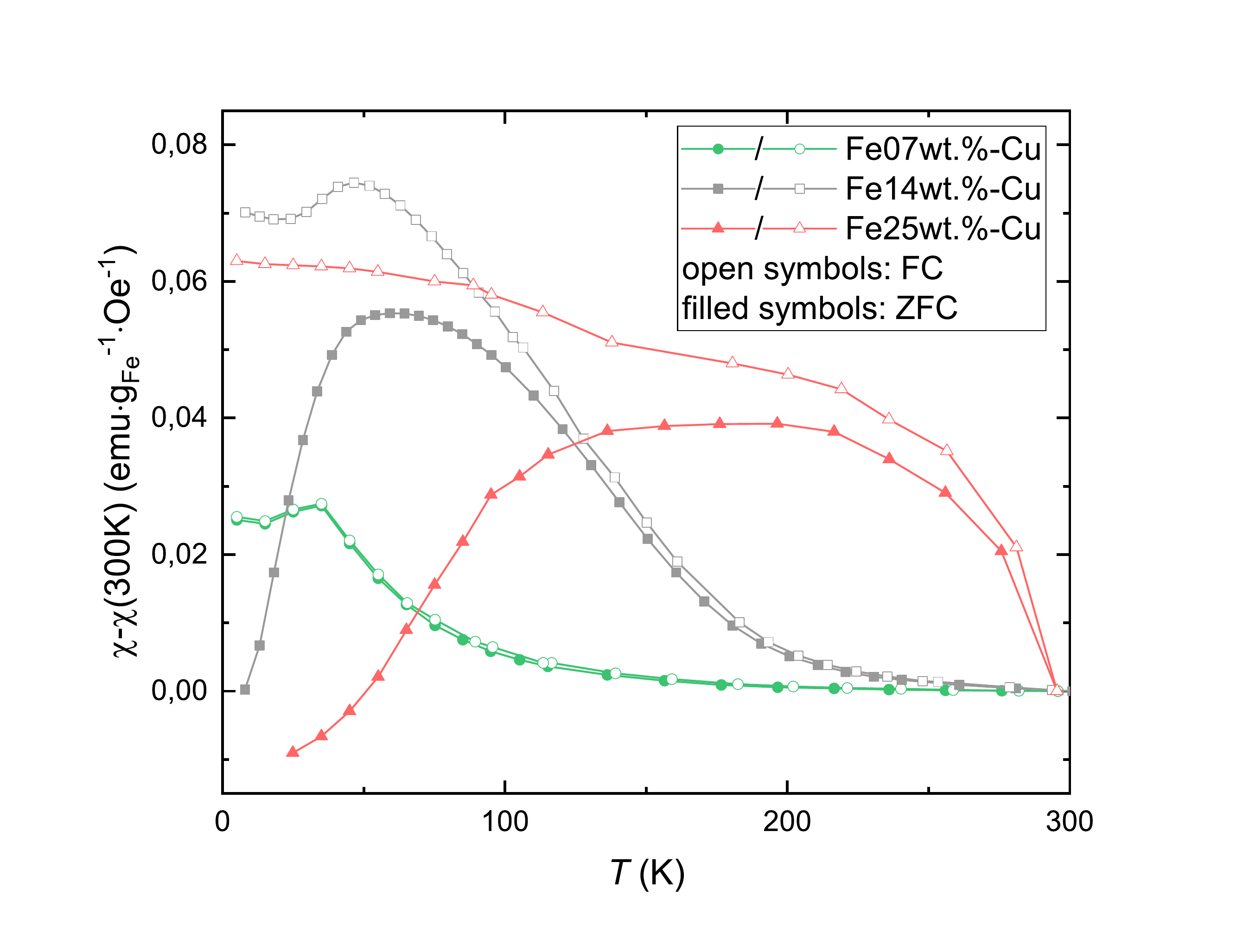}
\caption{Normalized zero-field cooling and field-cooling curves of as-deformed samples as a function of Fe-concentration. In all measurements a field of $H$=50~Oe was applied. Open symbols represent FC-curves, whereas filled symbols represent ZFC-curves.}
\label{fig:zfc_fc}
\end{figure}

The magnetization versus the applied magnetic field is measured in the as-deformed state for all compositions at 300~K and at 8~K (Fig.~\ref{fig:hyst_conc}). The hysteresis, measured at 300~K, do not saturate, even at the highest applied field of 70~kOe. The observed high-field susceptibility is expected to arise from diluted Fe with suppressed long-range interaction. A non-zero high-field susceptibility was also observed in measurements at 8~K. Although the hysteresis show paramagnetic features, non-zero coercivities are observed. Fig.~\ref{fig:coercivity} shows the coercivites deduced from hysteresis measurements for all investigated configurations. The coercivity as a function of annealing temperature shows the same development in all investigated compositions. This behaviour can be correlated with the microstructural states in this system. It is known, that HPT-processed samples exhibit a high amount of residual stresses $\sigma$ in the as-deformed state \cite{todt2018gradient}. The high magnetostriction $\lambda_S$ of Fe gives then rise to an enhancement of anisotropy by magnetoelastic anisotropy, which reads $K_{\sigma}{=}3/2\cdot\sigma\lambda_S$. Approximating the coercivity with the anisotropy field, as proposed in the Stoner-Wohlfarth model, $H_C{\propto}K_{eff}/M_S$, this further leads to an increase in coercivity \cite{shen2005soft}.  Upon annealing at 150\degree C neither grain growth nor demixing was observed, but recovery mechanisms lead to a reduction in residual stresses \cite{bachmaier2012formation}, which therefore lowers the coercivity. The coercivity of the as-deformed, as well as of the 150\degree C-annealed sample arises therefore from random anisotropy \cite{herzer1995soft}. Annealing at higher temperatures leads to demixing and grain growth. The coercivity as a function of annealing temperature increases for the 250\degree C-annealed sample to around 400~Oe. The random anisotropy model peaks at $d{=}\sqrt{A/K}$ and breaks down for larger grain sizes $d$. Using typical values for bcc-Fe \cite{kronmuller2003micromagnetism}, the peak coercivity can be calculated to $H_C{\approx}$600~Oe, in the order of the measured value. Annealing at 500\degree C leads to a drop in coercivity, indicating a crossover from single-domain to multi-domain behaviour with a classical $1/d$-scaling law. Fig.\ref{fig:ebsd} shows the phase map derived from an EBSD scan of the 500\degree C-annealed sample (Fe14wt\%-Cu). The Fe grain size was analyzed to $d$=(210$\pm$80)~nm. The corresponding coercivity can be calculated, leading to $H_C^{th}$=(100$\pm$50)~Oe, higher than the coercivity determined from hysteresis measurements ($H_C$=60~Oe). For HPT-processed samples, grains elongate in radial direction, therefore the grain size is underestimated in the direction observed in the EBSD scan, which might explain the deviation.
\\Additionally, low-field magnetic measurements are carried out. In Zero-Field Cooling (ZFC) measurements, the demagnetized sample is cooled in the absence of an external field. Starting from the lowest observed temperature, the magnetic moment is recorded during heating. In Field-Cooling (FC) measurements, the magnetic moment is recorded during cooling in an external field. Fig.~\ref{fig:zfc_fc} shows ZFC/FC-measurements for the as-deformed states.  For reasons of comparability, the data are normalized, i.e. the susceptibility at 300~K is subtracted from all values. Herein, Fe25wt.\%-Cu shows a large splitting and a broad peak in the ZFC-curve, which is a typical behaviour for thermal activation\cite{knobel2008superparamagnetism}. The ZFC/FC-curves of Fe14wt.\%-Cu also shows splitting of both curves, but in the FC-curve also a maximum at around 50~K can be identified. A local maximum in  the FC-curve is also observed in Fe07wt.\%-Cu. In the literature this feature is often attributed to a spin-glass behaviour of diluted Fe \cite{schneeweiss2017magnetic, chien1986magnetic, de2013nanoparticle}.
\subsection{Atom probe tomography}
\begin{figure*}
\includegraphics[width=0.3\linewidth]{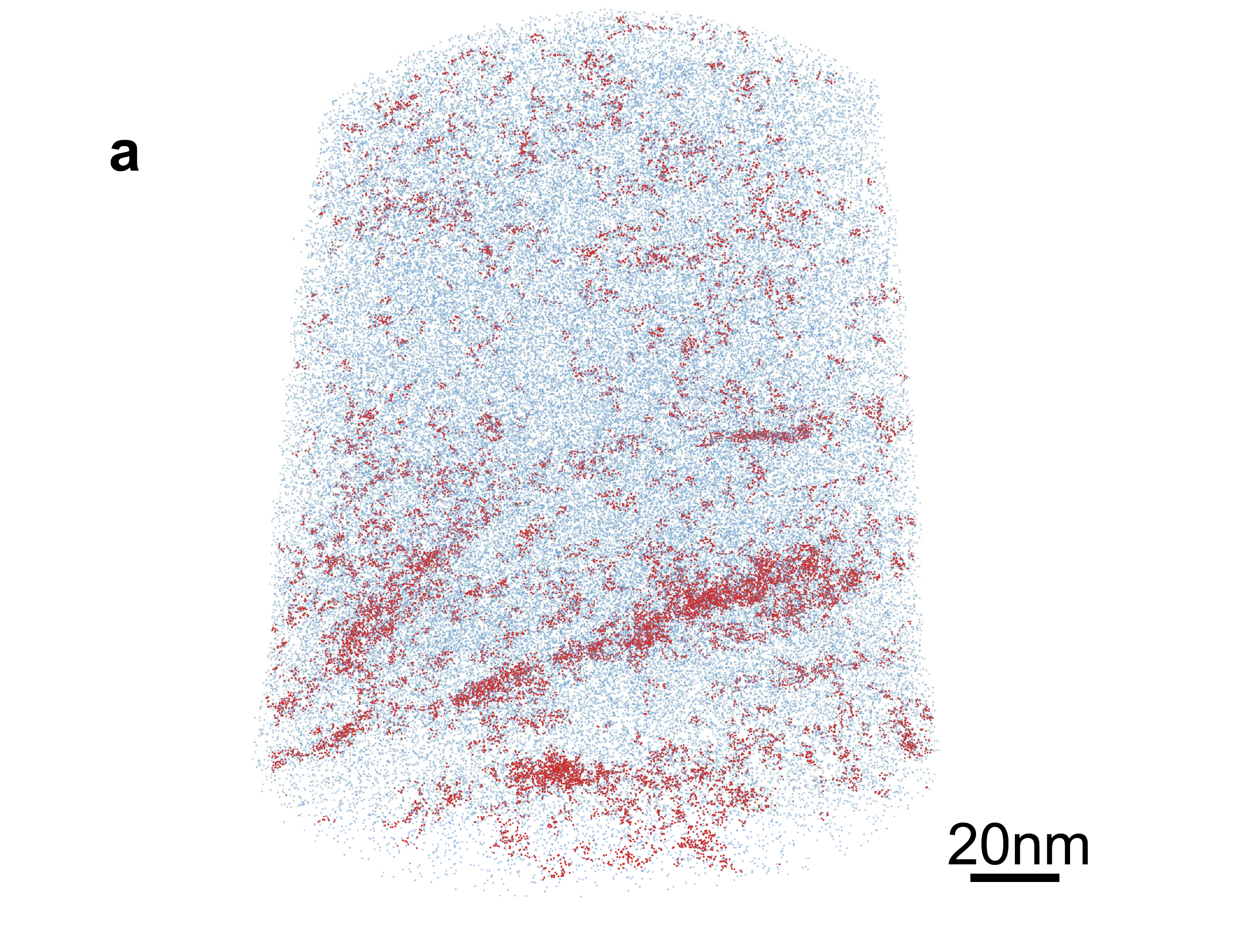}
\includegraphics[width=0.3\linewidth]{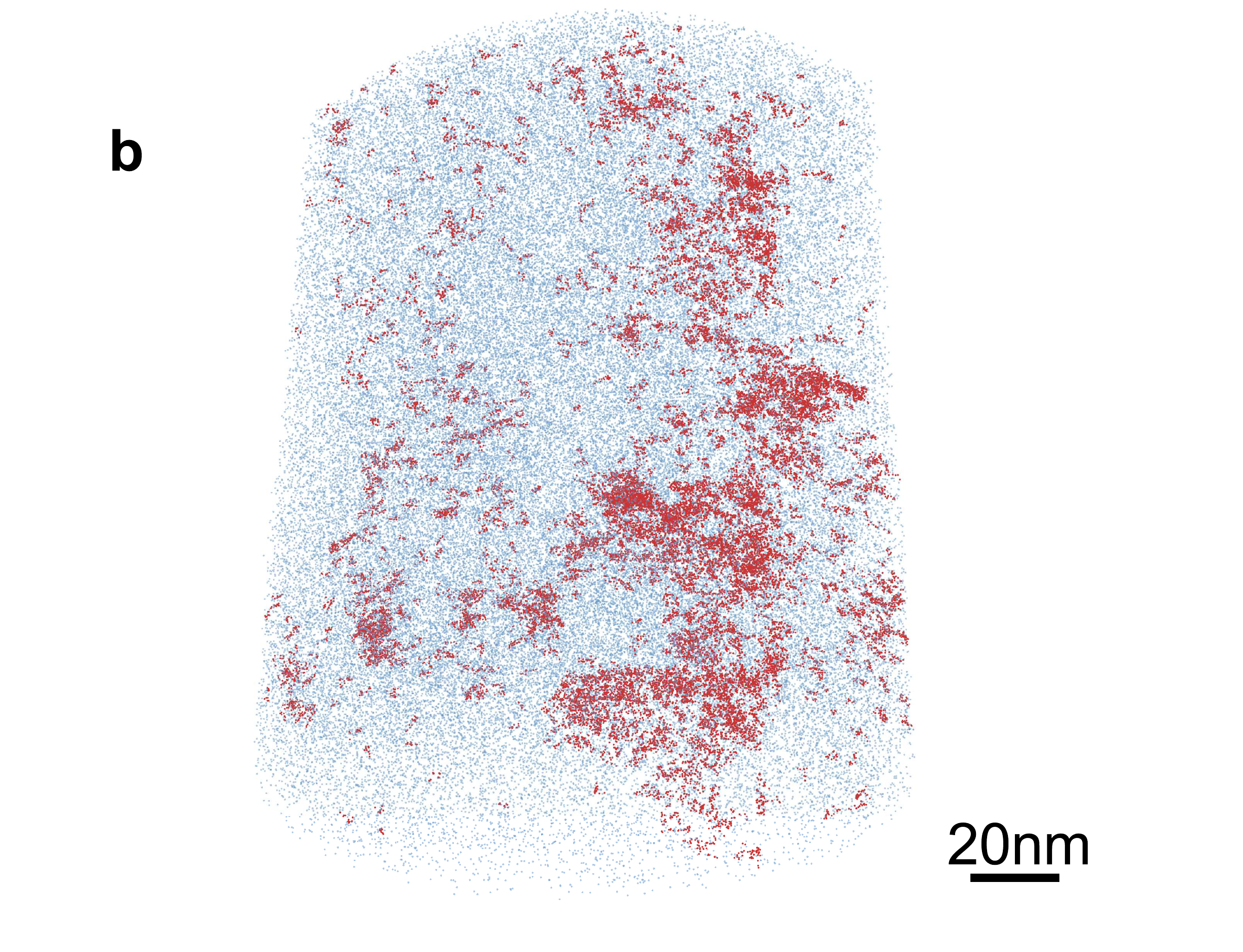}
\includegraphics[width=0.3\linewidth]{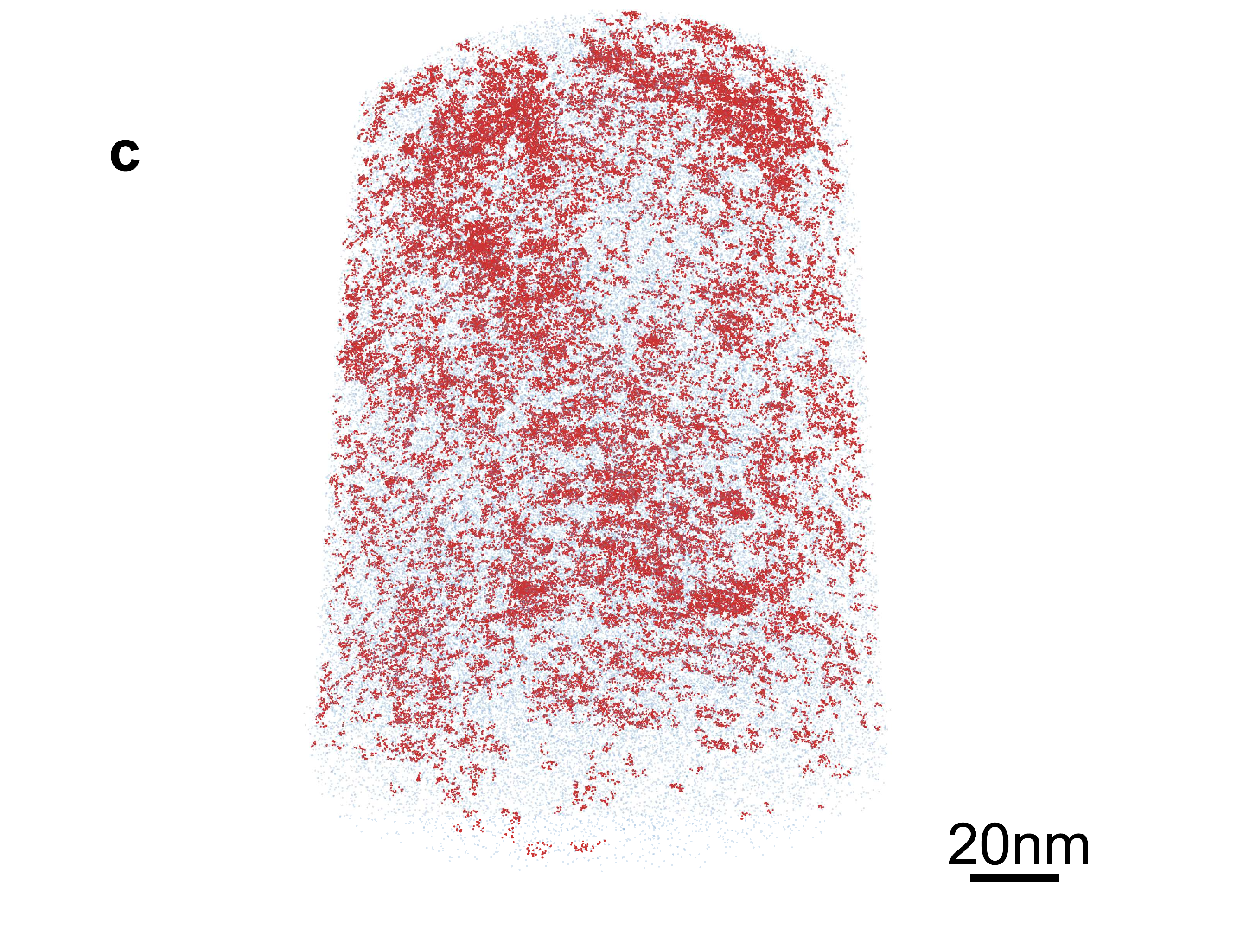}
\includegraphics[width=0.3\linewidth]{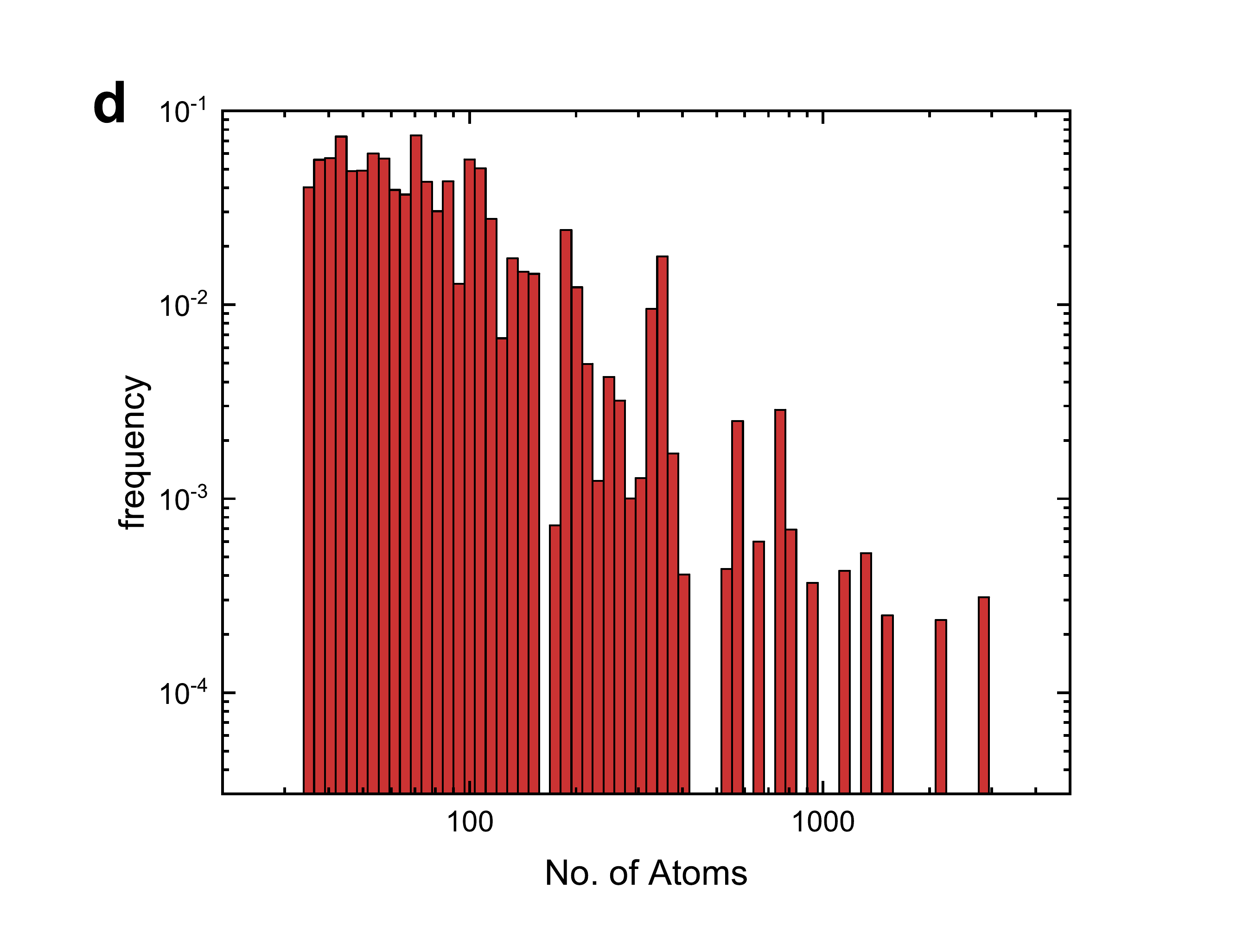}
\includegraphics[width=0.3\linewidth]{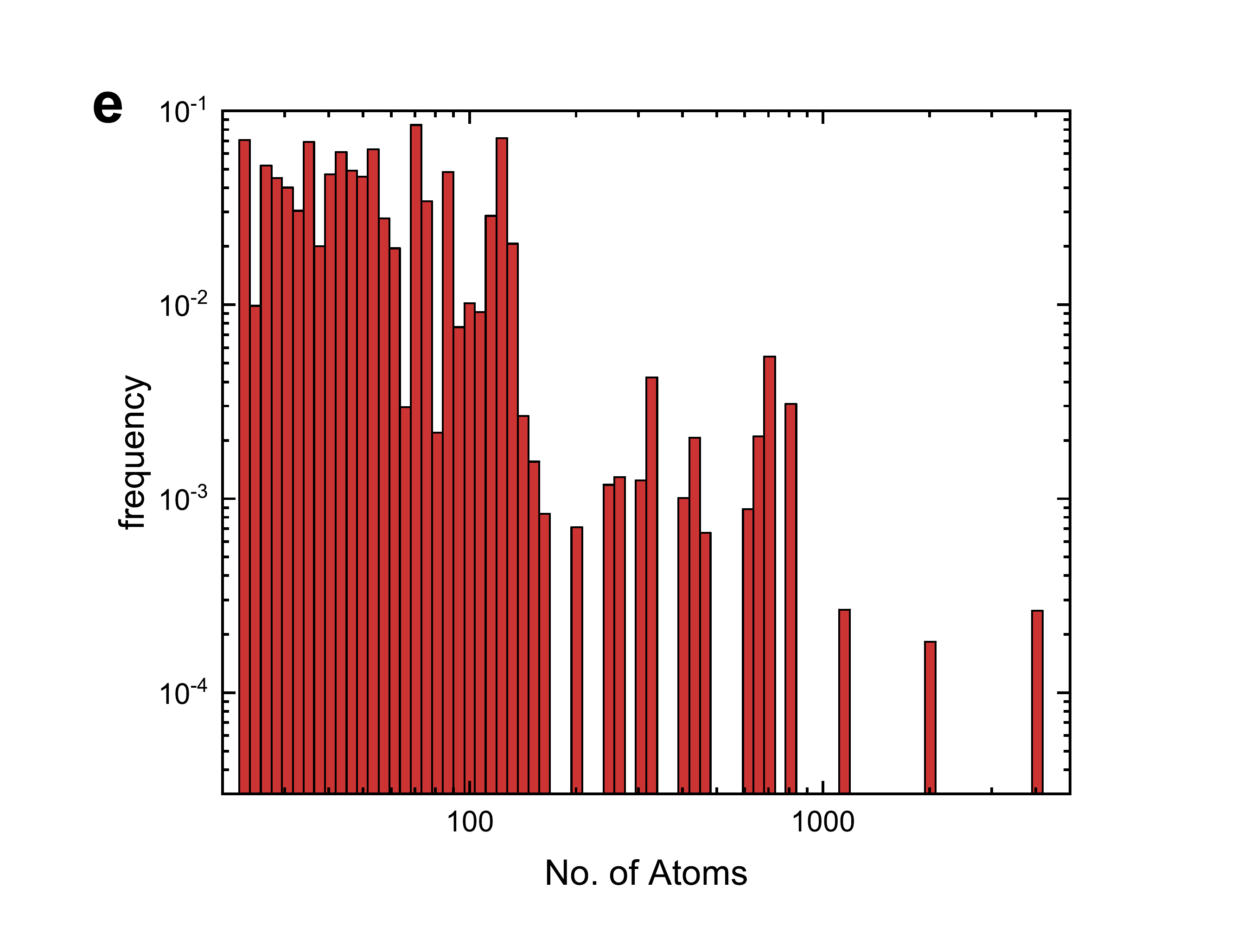}
\includegraphics[width=0.3\linewidth]{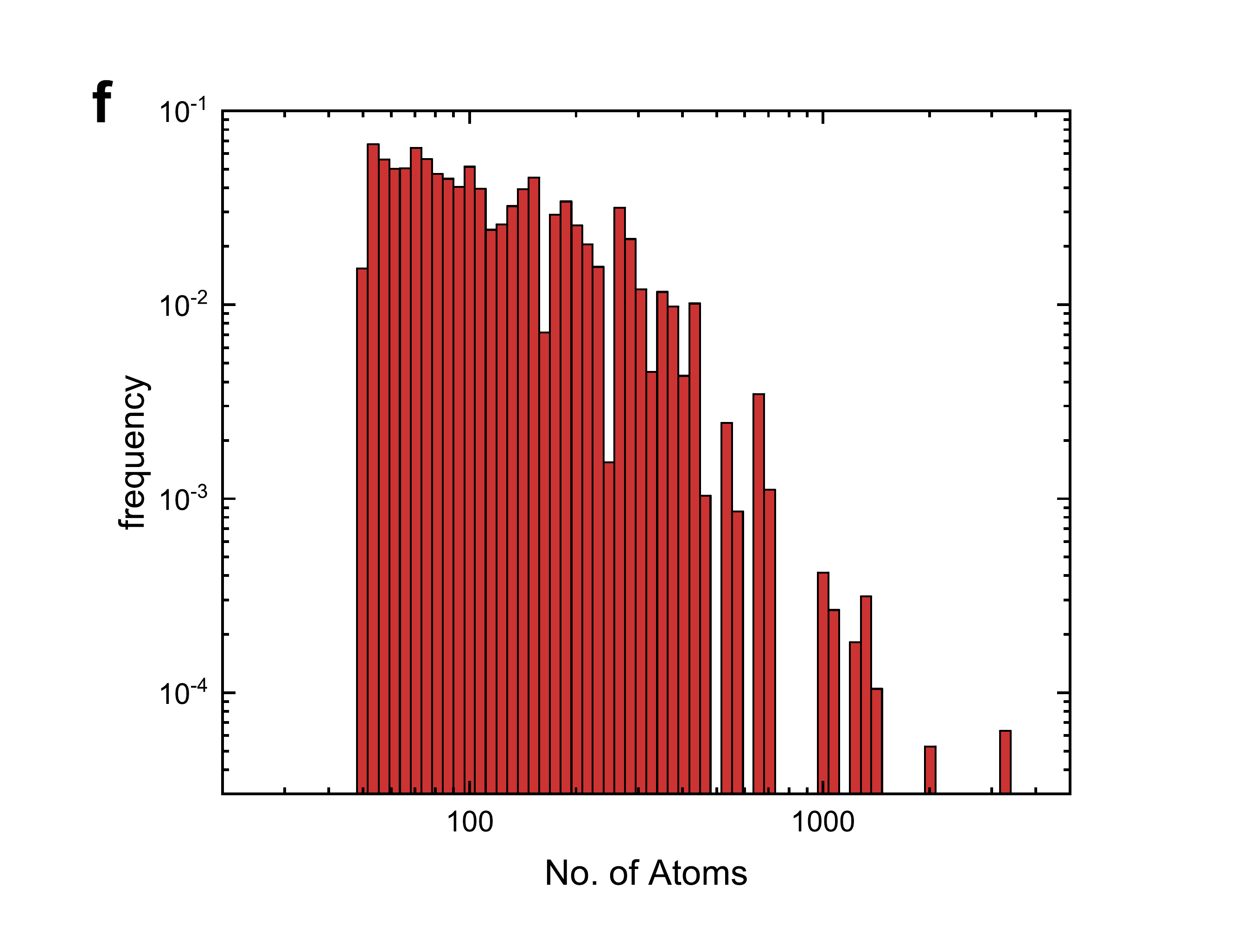}
\caption{APT reconstruction data for Fe07wt.\%-Cu (\textbf{a}), Fe14wt.\%-Cu (\textbf{b}) and Fe25wt.\%-Cu (\textbf{c}). 10$^5$ Fe-Atoms are plotted (blue dots). Red dots represent clustered Fe-Atoms. (\textbf{d})-(\textbf{f}) show the Fe-cluster size distributions of Fe07wt.\%-Cu, Fe14wt.\%-Cu and Fe25wt.\%-Cu, respectively.}
\label{fig:apt}
\end{figure*}

To investigate the distribution of Fe on the atomic scale, APT experiments are carried out. As mentioned above, the splitting in the ZFC/FC-curve is attributed to thermal activation, and therefore to the presence of Fe particles, small enough for thermal relaxation. Thus, the measured APT data is analyzed with respect to Fe-clusters, using Delaunay triangulation. Details on the used clustering algorithm can be found elsewhere \cite{felfer2015detecting}. In Fig.~\ref{fig:apt}(\textbf{a})-(\textbf{c}) the atom probe reconstruction of Fe07wt.\%-Cu, Fe14wt.\%-Cu and Fe25wt.\%-Cu respectively is plotted, showing the distribution of Fe. Therein, Fe25wt.\%-Cu shows large volumes of clustered Fe-atoms, as expected from the observed ZFC/FC-curve. In Fig.~\ref{fig:apt}(\textbf{d})-(\textbf{f}) the results from the clustering algorithm are shown. Qualitative differences arise, as the formation of Fe-clusters depends on the present local crystallographic defects. The cluster sizes span over a wide range, whereas the cluster size for Fe25wt.\%-Cu shows the broadest distribution. This can also be attributed to the observed maxima in the ZFC/FC-measurements (Fig.~\ref{fig:zfc_fc}), which are expected to arise from magnetic blocking. Anyway, in every sample about 95\% of all Fe-atoms are rejected by the clustering algorithm, meaning the majority of Fe-atoms is diluted in the Cu-matrix, which explains the large high-field susceptibility in Fig.~\ref{fig:hyst_conc}. One important parameter to achieve accurate spatial reconstruction is the detection efficiency of the used APT system. In \cite{stephenson2011estimating} it has been shown, that the cluster size distribution is systematically underestimated by the detection efficiency.

\section{Conclusion}
In this study, the magnetic and the microstructural properties of HPT-processed binary Fe-Cu samples, in the as-deformed as well as in annealed states, are correlated, whereas a specific focus is on the investigation of the amount of intermixing. In the as-deformed states, Fe is present in two different configurations, namely diluted in the fcc-Cu matrix, as well as in form of clusters. The ratio of clustered and diluted Fe varies as a function of composition and gives rise to either thermal relaxation (in case of Fe25wt.\%-Cu), magnetic frustration (in case of Fe07wt.\%-Cu) or a superposition of both, which is true for Fe14wt.\%-Cu. The results further prove that intermixing of Fe and Cu can be obtained with HPT on an atomic scale. The coercivity as a function of the annealing temperature complies the random anisotropy model, showing Fe-particle sizes below the exchange length persist in annealed states, which further gives rise to magnetic tunability.

\section{acknowledgments}

This project has received funding from the European Research Council (ERC) under the European Union’s Horizon 2020 research and innovation programme (Grant No. 757333). Synchrotron measurements have been performed at PETRA III: P07 at DESY Hamburg (Germany), a member of the Helmholtz Association (HGF). We gratefully acknowledge the assistance by N. Schell. The authors thank F. Spieckermann and C. Gammer for their help with data processing.

\bibliography{magndil}

\end{document}